\documentclass[aps, prd, superscriptaddress, nofootinbib, preprintnumbers 
]{revtex4}
\usepackage{amsmath}
\usepackage{graphicx}
\usepackage{color}

\newcommand{\lesssim}{\mathrel{\mathpalette\vereq<}}

\newcommand{\chushi}[1]{}

\definecolor{mygrn}{rgb}{0,0.5,0}
  
\begin{document}

\preprint{EPHOU-16-020}

\title{Bosonic-Seesaw Portal Dark Matter}      
\author{Hiroyuki Ishida}\thanks{{\tt hiroyuki403@cts.nthu.edu.tw}}
	\affiliation{ Physics Division, National Center for Theoretical Sciences, Hsinchu 30013, Taiwan.}
\author{Shinya Matsuzaki}\thanks{{\tt synya@hken.phys.nagoya-u.ac.jp}}
      \affiliation{ Institute for Advanced Research, Nagoya University, Nagoya 464-8602, Japan.}
      \affiliation{ Department of Physics, Nagoya University, Nagoya 464-8602, Japan.}   
\author{Yuya Yamaguchi}\thanks{{\tt yy@particle.sci.hokudai.ac.jp}}
	\affiliation{Department of Physics, Faculty of Science, Hokkaido University,
 Sapporo 060-0810, Japan.}


\begin{abstract}
We discuss a new type of Higgs-portal dark matter (DM)-production
mechanism, called bosonic-seesaw portal (BSP) scenario.
The BS provides the dynamical origin of the electroweak
symmetry breaking, triggered by mixing between
the elementary Higgs and a composite Higgs generated by
a new-color strong dynamics, hypercolor (HC).  
At the HC strong coupling scale, 
the classical-scale invariance assumed in the model 
is dynamically broken as well as the ``chiral'' symmetry present in the HC sector. 
In addition to the composite Higgs, 
HC baryons emerge to potentially be stable 
because of the unbroken HC baryon number symmetry.    
Hence the lightest HC baryon can be a DM candidate.  
Of interest in the present scenario is that 
HC pions can be as heavy as the HC baryon due to 
the possibly enhanced-explicit ``chiral''-breaking effect 
triggered after the BS mechanism, so the HC baryon 
pair cannot annihilate into HC pions. 
As in the standard setup of freeze-in scenario, 
it is assumed that the DM was never in the thermal equilibrium, 
which ends up with no thermal abundance.  
It is then the non-thermal BSP process that crucially 
comes into the game below the HC scale: 
the HC baryon significantly couples to the standard-model Higgs via the 
BS mechanism, and can non-thermally be produced from the thermal plasma 
below the HC scale, which turns out to 
allow the TeV mass scale for the composite baryonic DM, 
much smaller than the generic bound placed in the conventional thermal 
freeze-out scenario, to account for the observed relic abundance.  
Thus the DM can closely be related to the mechanism of the 
electroweak symmetry breaking. 
\end{abstract}
\maketitle

\section{Introduction} 
Quarter of our universe is constituted by unknown matter, called the dark matter (DM). 
Several cosmological and astrophysical observations have so far suggested that  
the DM should be electrically neutral, cold enough, and stable enough to be long-lived compared 
with the age of the universe. 
The abundance left in the present universe is thought to have been produced 
via interactions with the standard model (SM) particles in the early universe, involving 
called mediators such as the SM Higgs (Higgs portal scenario)~\cite{Silveira:1985rk}.

In this paper, we discuss a new type of the Higgs-portal DM-production mechanism, 
which we call the bosonic-seesaw portal (BSP) scenario. 
The DM candidate will be identified as a composite baryonic state, 
arising as a bound state of new fermions 
strongly coupled in a new-color dynamics (hypercolor (HC)). 
The HC triggers the electroweak symmetry breaking (EWSB) via the 
BS mechanism~\cite{Antipin:2014qva,Haba:2015qbz,Ishida:2016ogu}, 
which also generates the portal coupling between  
the DM candidate and the 125 GeV Higgs boson, dubbed as the BSP coupling.  
It is the BSP process that 
produces the DM relic abundance, the right amount of which 
can be achieved in accordance with the BS mechanism.

Composite baryonic dark matters 
have been discussed in Refs.~\cite{Antipin:2014qva,Griest:1989wd,Blum:2014dca}. 
The DM abundance in those studies is 
thermally produced through the annihilation to the HC pions, 
where, by a naive analogy to QCD, 
the thermal relic abundance can be estimated as 
$\Omega_{\rm DM} h^2 \simeq 10^{-5}$ 
when we suppose the order of TeV mass for the DM.  
This implies that, 
to realize the correct amount of the presently observed DM abundance, 
a general limit on the composite baryonic DM mass 
is set by the thermal freeze-out 
relic abundance, $m_{\rm DM} = {\cal O}(100)$ TeV~\cite{Antipin:2014qva,Griest:1989wd,Blum:2014dca}.

In the scenario addressed here, 
in contrast to the conventional freeze-out scenario, 
the DM abundance is not thermally produced 
from the strong HC sector itself.  
The production of the DM takes place 
non-thermally by the BSP process (which is {\it \`{a} la} freeze-in scenario~\cite{Blennow:2013jba}), 
outside of the HC sector. 
It then turns out that the BSP production 
allows the DM mass to be on the order of TeV. 
Thus the relic abundance of somewhat light composite 
DM directly links to the EWSB mechanism.  

The present scenario would generically be operative as long as 
the following two setups are at hand:  

\begin{itemize} 

\item[{\bf (I)}] 

the classical scale invariance is present 
to be dynamically broken by the HC sector  
weakly coupled to the SM sector, 
yielding the BS mechanism.

\item[{\bf (II)}]

 HC pions get massive after the BS mechanism (i.e. below the 
strong HC coupling scale) due to the weak coupling to  
another hidden sector, which dynamically enhances 
the explicit breaking effect of ``chiral'' symmetry 
present in the HC sector, to make the HC pion mass as large as 
the HC dynamical scale, i.e. HC baryon mass scale 
(DM mass scale). 
In this setup, the HC baryon cannot annihilate into HC pions 
and the inclusive annihilation cross section would  
yield no thermal relic abundance (which would be 
the initial condition for the DM thermal history).

\end{itemize}

To demonstrate the point, 
as a concrete example we shall take a BS model discussed 
in Refs.~\cite{Haba:2015qbz,Ishida:2016ogu}.

\section{Scenario description: part (I)}

We begin by employing the former two sectors in the above part {\bf (I)}, 
which are constructed from the classically scale invariant SM 
and the HC gauge sector of $SU(N_{HC})$. 
In the model we have the HC gluon as well as 
the HC fermions $F_{L/R}$ forming 
the $SU(3)_{F_{L/R}}$-flavor triplet, 
$F_{L/R} = (\chi_{L/R},\psi_{L/R})^T$, where $\chi$ denotes the $SU(2)_{F_{L/R}}$ 
doublet.   
These HC fermions carry the vector-like charges under 
the $SU(3)_c \times SU(2)_W \times U(1)_Y \times SU(N_{\rm HC})$ gauge groups, 
$\psi_{L/R}\sim (1, 1, 0, N_{\rm HC})$ 
and $\chi_{L/R} \sim (1, 2, 1/2, N_{\rm HC})$. 
The HC fermions couple to the elementary Higgs doublet $H$ 
with the small coupling $y$ as 
\begin{equation} 
- y \cdot 
\bar{F}_L 
\left( 
\begin{array}{cc} 
{\bf 0}_{2 \times 2} & H \\ 
H^\dag & 0 
\end{array} 
\right) 
F_R 
+ {\rm h.c.} 
= 
- y \cdot \bar{\chi} H \psi + {\rm h.c.} 
\,.  \label{y}
\end{equation} 
At around the scale $\Lambda_{\rm HC}$, 
the HC gauge coupling gets so strong that     
the ``chiral'' $SU(3)_{F_L} \times SU(3)_{F_R}$ 
symmetry is spontaneously broken by the ``chiral''(but SM vector-like) 
condensate $\langle \bar{F}^aF^b \rangle \sim - 4 \pi f_{\pi_{\rm HC}}^3/\sqrt{N_{\rm HC}} 
\delta^{ab}$ ($a,b=1,\cdots, 8$), 
down to the vectorial $SU(3)_{F_V}$, 
where $f_{\pi_{\rm HC}} \sim \sqrt{N_{\rm HC}} \cdot 
\Lambda_{\rm HC}/(4 \pi)$ is the HC pion 
decay constant.   
After the HC confinement, furthermore, 
the HC fermions form composite HC hadrons  
in a manner similar to the QCD case. 
Among those HC hadrons,  
a composite HC scalar doublet $\Theta \sim \bar{\psi} \chi$ 
($\Theta^\dag \sim \bar{\chi} \psi$), 
embedded in the $SU(3)_F$ flavor nonet,  
has the same quantum 
numbers as those the elementary Higgs doublet $H$.  
Hence this $\Theta$ mixes with the elementary Higgs doublet 
$H$, due to the presence of the above Yukawa term, 
so that one finds the mixing form 
\begin{equation} 
- y \cdot f_\Theta m_\Theta   
(H^\dagger \Theta) + {\rm h.c.}
\,, 
\end{equation} 
where $m_\Theta$ denotes the mass of composite Higgs doublet $\Theta$, 
which is of ${\cal O}(\Lambda_{\rm HC})$~\footnote{
The composite Higgs doublet $\Theta$ should not be confused with 
pseudo Nambu-Goldstone bosons as in composite Higgs models in the market: 
all ``chiral'' HC pions in the present scenario are CP-odd pseudoscalars 
($\pi_{\rm HC}^{ab} \sim \bar{F}^a i \gamma_5 F^b$ with $J^{P}=0^{-}$), 
not CP-even scalars.},  
and $f_\Theta$ is the decay constant associated with the scalar current 
$\bar{\chi}\psi$ coupled to the $\Theta$, 
defined as $\langle 0 | \bar{\chi} \psi(0) | \Theta \rangle = f_\Theta m_\Theta$.  
The induced mixing term yields 
the mass matrix for $(H, \Theta)^T$:  
\begin{equation} 
  \left(
  \begin{array}{cc}
 0 & y \Lambda_{\rm HC}^2\\ 
 y \Lambda_{\rm HC}^2 &  \Lambda_{\rm HC}^2 
  \end{array}
\right)
\,, \label{mixing-matrix}
\end{equation} 
with the mixing strength $y \ll 1$ controlling the coupling 
between the SM and the HC sectors.  
In Eq.(\ref{mixing-matrix}) 
we have simply taken $f_\Theta \simeq m_\Theta \simeq \Lambda_{\rm HC}$, 
which can be expected from the QCD case~\footnote{
In terms of the $SU(3)$ flavor nonet in QCD, the $\Theta$ can be 
viewed as an analogue of the scalar meson $K^*_0(1430)$ with 
$I(J^P)=\frac{1}{2}(0^+)$. 
}.  
Note that the determinant of the mass matrix is negative, so the negative mass-squared of the SM Higgs 
is dynamically generated   
by the seesaw mechanism (the BS mechanism~\cite{BSS}) 
to trigger the EWSB (For detailed potential analysis, see  Ref.~\cite{Ishida:2016ogu}). 

In addition to the composite Higgs, 
the HC sector generically involves the rich composite spectra
such as the HC pions and baryons as in the case of QCD. 
Among those HC hadrons, 
some of HC baryons, presumably the lightest one 
can be stable due to the conserved HC baryon number, 
hence can be a DM  candidate~\footnote{
The HC baryon number associated with the unbroken $U(1)_{F_V}$ is necessarily conserved, as long as 
the HC dynamics is vectorlike and the HC fermions are vector-likely charged as in the present model. 
}.  
Here we shall suppose that the HC sector possesses $N_{\rm HC}=4$, i.e. 
the $SU(4)_{\rm HC}$. 
The HC baryon can be realized as a complex scalar, 
having the HC scalar-baryon charge~\footnote{In Refs.~\cite{LSD}, a similar composite dark-bosonic baryon as the DM candidate, 
called stealth DM, 
has been discussed in a context different than the bosonic seesaw.}.  
(Our argument is substantially unchanged even if we employ the case other than the $SU(4)_{\rm HC}$ in which 
fermionic HC baryons can be present, as will be clarified below.)  
Among the HC baryons, the EW-singlet HC scalar-baryon, $\varphi \sim \psi \psi \psi \psi$ 
can be the lightest, i.e., the DM candidate. 
The $\varphi$ mass is expected to be on the order of 
${\cal O}((N_{\rm HC}/3) \Lambda_{\rm HC})$. 
In the present article, we take $\Lambda_{\rm HC}={\cal O}$(TeV), 
hence $m_\varphi ={\cal O}$(TeV) as well.

The DM candidate $\varphi$, the EW-singlet complex scalar baryon, strongly and minimally
 couples to the composite HC Higgs doublet, $\Theta$, like 
\begin{equation} 
 a \cdot \varphi^\dag \varphi \Theta^\dag \Theta 
 \,, 
\end{equation}
 with the order one (or larger) coefficient $a$~\footnote{ 
The coefficient $a$ does not scale with $N_{\rm HC}$, i.e., $a={\cal O}(N_{\rm HC}^0)$, 
because it is the coupling of the baryon-meson two-body scattering.}.  
By the BS mechanism, 
the $\Theta$ starts to mix with the elementary Higgs doublet $H$ 
below the scale $\Lambda_{\rm HC}$. 
This dynamically generates a Higgs portal coupling between the DM $\varphi$ 
and the SM Higgs $H_1$: 
\begin{equation} 
 \kappa_{\varphi H} \cdot \varphi^\dag \varphi H_1^\dag H_1 
 \,, \qquad {\rm with} 
 \qquad 
 \kappa_{\varphi H} = a y^2 
\,,  \label{Higgs-portal}
\end{equation}
where the factor $y^2$ has come from 
the BS mixing strength $y$ ($\Theta \approx y H_1 + H_2$, which can be understood by 
diagonalizing the mass matrix Eq.(\ref{mixing-matrix}))  
between the SM Higgs $H_1$ and heavy Higgs $H_2$.  
The mixing strength $y$ is supposed to be 
much smaller than ${\cal O}(1)$, 
so that the Higgs portal coupling  $\kappa_{\varphi H}$ 
can naturally be  small to be consistent with the present relic 
abundance of the dark matter, as will be clarified later on.

\section{Scenario description: part (II)}

The spontaneous breaking of the 
``chiral'' $SU(3)_{F_L} \times SU(3)_{F_R}$ symmetry gives rise to 
8 Nambu-Goldstone bosons, HC pions $(\pi_{\rm HC})$. 
The ``chiral" symmetry is explicitly broken by 
the $y$-Yukawa interaction in Eq.(\ref{y}), 
which makes HC pions in part massive with 
the tiny mass, $\Delta m_{\pi_{\rm HC}}^2|_{y} \sim {\cal O}(y \cdot v \Lambda_{\rm HC})$, where 
$v$ is the EW scale $\simeq 246$ GeV ($={\cal O}(\Lambda_{\rm HC}/5)$). 
Also the EW gauge interactions would slightly lift up 
the masses for a part of HC pions charged under the EW, 
$\Delta m_{\pi_{\rm HC}}^2|_{\rm EW} \sim {\cal O}(\alpha_{\rm em} \Lambda_{\rm HC}^2)$.  
Note that by those explicit breaking effects, 
the HC pions cannot have the mass as large as the HC scale. 
To make the HC pions heavy enough by explicit breaking outside of the HC and EW dynamics,   
thus one needs another hidden sector explicitly breaking the ``chiral" symmetry, 
as described in the part {\bf (II)} in the Introduction.

In the BS model proposed in Refs.~\cite{Haba:2015qbz,Ishida:2016ogu}, 
such an explicit breaking term is supplied by 
a pseudoscalar ($S$) in the scale-invariant form as 
\begin{equation} 
 g_S \cdot (\bar{F} i \gamma_5 F) S 
\,, 
\end{equation} 
with the weak coupling $g_S(\ll 1)$, which ensures conservation of 
the approximate ``chiral" symmetry in the theory. 
Below the HC confinement scale $\sim \Lambda_{\rm HC}$, 
the pseudoscalar $S$ dynamically develops its vacuum expectation value, 
$v_S$,  
due to  
the pseudoscalar seesaw mechanism between $S$ and 
the HC $\eta'$ meson~\cite{Haba:2015qbz,Ishida:2016ogu},  
and makes all the HC pions massive by the $g_S$-Yukawa coupling as~\cite{Ishida:2016ogu} 
\begin{equation} 
 m_{\pi_{\rm HC}} \sim \left( \frac{\Lambda_{\rm HC}}{f} \right) g_S v_S 
\,, 
\end{equation} 
where $f=f_{\pi_{\rm HC}}/\sqrt{N_{\rm HC}/3}(\sim \Lambda_{\rm HC}/(4\pi))$ 
with the HC pion decay constant $f_{\pi_{\rm HC}}$.   
Thus the large $v_S$ enhances the explicit breaking effect to 
make it possible to lift the HC pion mass to 
reach the scale as large as ${\cal O}(\Lambda_{\rm HC})$, i.e., the HC baryon $(\varphi)$ 
mass scale: 
\begin{equation} 
m_{\pi_{\rm hC}} \sim m_\varphi 
\sim {\cal O}(\Lambda_{\rm HC})
\,. \label{mpi}
\end{equation} 
Actually, as discussed in Ref.~\cite{Ishida:2016ogu}, 
the $v_S$ is required to be much larger than $\Lambda_{\rm HC}$ 
to realize the EWSB through the potential analysis, 
where the HC pion mass is indeed lifted up to be on the order of 
$\Lambda_{\rm HC}$, even the small $g_S$ coupling.  
Crucial to note here is that this explicit-breaking 
enhancement is triggered 
after the HC confinement/the ``chiral'' symmetry breaking (as well as 
the BS mechanism) takes place, 
above which scale the $g_S$ term does not play any role, so 
the HC fermions are (almost) massless~\footnote{
This point is also different from the heavy pseudoscalar in QCD, 
such as pseudoscalar mesons including charm and bottom quarks, 
which acquire the large mass because of the large current quark masses $m_c, m_b$. 
}. 
Hence this effect gets relevant only for the HC pions, 
not for HC fermions, or HC baryons, either.~\footnote{
The amplification of explicit breaking effect irrespective to bare fermion mass 
has been addressed in a different strong dynamics in \cite{Matsuzaki:2015sya}, and other references therein. }

Note also that the presence of the large $v_S$ is necessary to realize the  
EWSB scale in the present BS model, 
 as was demonstrated in  Ref.~\cite{Ishida:2016ogu}.

The heavy HC pion will significantly affect the thermal history of 
the lightest HC baryon $\varphi$:   
the $\varphi$ was generated when the temperature cools down to 
the critical scale of ${\cal O}(\Lambda_{\rm HC})$.   
Since the $\varphi$ mass is on the same order as $\Lambda_{\rm HC}$ 
analogously to the case of QCD light baryons (e.g. neutron and proton), 
the number density of $\varphi$ gets diluted no sooner than 
the HC confinement. 
Due to the possible breaking enhancement as in Eq.(\ref{mpi}), 
such a $\varphi$ cannot kinematically annihilate into HC pions, 
so that the conventional thermal freeze-out would not happen, 
which could drastically alter the composite baryonic DM scenario, 
as will be discussed in detail below~\footnote{ 
The heavy HC pions having the mass of ${\cal O}({\rm TeV})$ as in Eq.(\ref{mpi}), 
including electromagnetically neutral ones, 
completely decay to EW gauge bosons with 
the width of ${\cal O}(100)$ GeV~\cite{Ishida:2016ogu}, 
to quickly disappear below the HC scale $\Lambda_{\rm HC}$.}.

\section{Non-thermal DM production by BSP}

\begin{figure}[t]
\begin{center}
   \includegraphics[width=7.0cm]{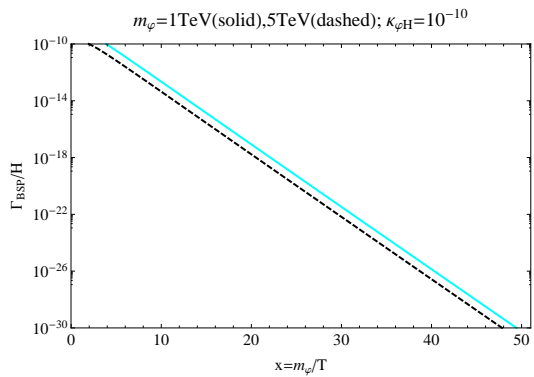} 
\caption{ 
The ratio of the BSP reaction rate to the Hubble parameter, 
$\Gamma_{\rm BSP}/H$, where $\Gamma_{\rm BSP}=n_\varphi \langle \sigma_{\rm BSP} v_{\rm rel} \rangle$ with the number density of the (non-relativistic and 
bosonic) DM $\varphi$ 
($n_\varphi \simeq 1/\pi^2 (m_\varphi T/2\pi)^{3/2} e^{- m_\varphi/T}$) 
and the 
thermal-averaged cross section estimated by summing the production 
cross sections in Eq.(\ref{sigmas}).  
In the plot we have assumed the radiation dominance for 
the Hubble parameter $H$ ($H \simeq 0.33 \sqrt{g_*(T)} T^2/M_p$) with the 
effective degree of freedom $g_*(T)$ set to 100, 
and taken the BSP coupling $\kappa_{\varphi H} = 10^{-10}$ 
consistently with the estimate in Eq.(\ref{upper}). 
\label{Gamma-H}} 
\end{center} 
\end{figure}

\begin{figure}[t]
\begin{center}
   \includegraphics[width=7.0cm]{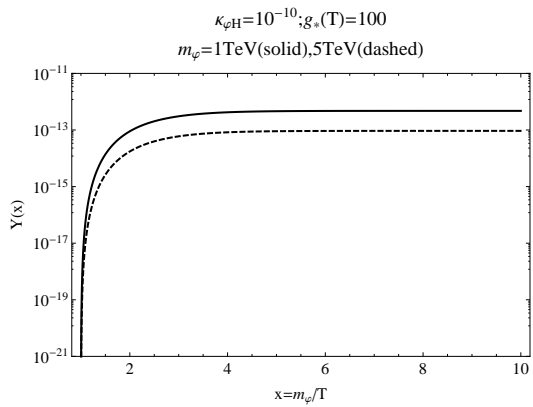} 
\caption{ 
A sample of the thermal history of the HC scalar baryon 
$\varphi$ yield (number density per entropy density) $Y=n/s$ 
against the temperature x=$m_\varphi/T$ along the bosonic seesaw portal production, 
with the mass $m_\varphi = 1$ TeV (solid curve), 5 TeV (dashed curve) 
and the coupling $\kappa_{\varphi H}=10^{-10}$ fixed 
(for $N_{\rm HC}=4$). 
The effective degree of freedom $g_*(T)$ has been set to 100 in the plot.       
\label{y-x}} 
\end{center} 
\end{figure}

Now we discuss the production mechanism 
for the composite baryonic DM based on the scenario descriptions {\bf (I)} and 
{\bf (II)}. Of importance to note is the presence of the Higgs 
portal coupling in Eq.(\ref{Higgs-portal}), dynamically induced from 
the BS mechanism at $T=\Lambda_{\rm HC}$.  
Although the number density of the $\varphi$ 
has been diluted, 
due to the induced portal coupling, 
the $\varphi$ can be produced 
unilaterally via the SM sector such as 
SM SM $\to$ $\varphi \varphi^\dag$.   
Note that, although the composite DM strongly couples in the HC sector, 
the BSP interaction  is perturbative due to the suppression by $y \ll 1$,  
hence can be computed reliably enough by the usual perturbation theory. 
It thus turns out that 
the BSP interaction is non-thermal for whole time scale in the thermal history 
and the produced 
abundance saturates slightly after the generation of $\varphi$: 
at around $x= m_\varphi/T \simeq \Lambda_{\rm HC}/T \simeq 5$,  
see Fig.~\ref{Gamma-H} and \ref{y-x}. 
(If the HC confinement would 
happen at somewhat lower temperature, say lower by factor of 10 than 
the HC baryon mass, $T_c \sim 1/10 \Lambda_{\rm HC}$, as in the case of QCD,  
the thermal history described in Fig.~\ref{y-x} would be changed just by  making  
the starting position of the rising up shifted to $x=m_\varphi/T \sim 10$~\footnote{
Then the HC and EW phase transitions might almost simultaneously happen in the thermal history. 
In that case the abundance estimate for the BSP process produced from SM particles could be more complicated than that done in the present analysis.}.) 

As noted in the previous section, 
the HC pions become heavy enough such that the DM population is 
not sufficiently created by the heavy HC pions 
and the DM annihilation into two HC pions is kinematically blocked.   
Even that case, however, 
the thermal DM abundance produced by the total inclusive annihilation cross section 
might not be negligibly small because 
of the strong interaction of the HC sector.  
Then 
the thermal history of the DM would be changed from the one in Fig.~\ref{y-x} 
and reduced to be a conventional freeze-out scenario, where 
the freeze-out takes place at around $x\sim 20-30$, which is much later than 
the BSP production at $x\sim 5$.  
To separate our present scenario from 
the conventional freeze-out scenario, 
we shall therefore assume that the DM was never in the thermal equilibrium, 
as done in the usual freeze-in scenario as a natural setting~\cite{Blennow:2013jba}.   
With this assumption at hand, we may then 
safely set the initial condition for 
the number density per comoving volume, $Y(T=m_\varphi) = 0$. 

We thus evaluate the production cross sections 
arising from the BSP coupling in Eq.(\ref{Higgs-portal}). 
Since the $\varphi$ is generated at 
 $T=\Lambda_{\rm HC}$ and the saturation point $(T_f)$ at around the EW scale 
($T_f=m_\varphi/5 ={\cal O}(v)$, see Fig.~\ref{y-x}),   
the relevant processes governed by the population in the thermal bath are:  
$hh, f\bar{f}, WW, ZZ \to \varphi^\dag \varphi $. 
Those cross sections are computed at the tree-level of the perturbation in couplings 
to be 
\begin{eqnarray} 
\sigma(hh \to \varphi^\dag \varphi) 
&=& 
\frac{\kappa_{\varphi H}^2}{16 \pi} \frac{s}{(s - m_h^2)^2} 
 \frac{\left(1 - \frac{4 m_\varphi^2}{s} \right)^{5/2}}{\sqrt{1-\frac{4 m_h^2}{s}}} 
 \,, \nonumber \\ 
 \sigma(WW/ZZ \to \varphi^\dag \varphi) 
&=& 
\frac{9 \kappa_{\varphi H}^2}{64 \pi} \frac{s}{(s - m_h^2)^2} 
 \frac{\sqrt{1 - \frac{4 m_\varphi^2}{s}}}{\sqrt{1-\frac{4 m_{W/Z}^2}{s}}}   
\left( \frac{m_{W/Z}^2}{s}  \right)^2 
\left[2 + \left(  1 - \frac{s}{2 m_{W/Z}^2} \right)^2 \right] 
\,, \nonumber \\ 
\sigma(f \bar{f} \to \varphi^\dag \varphi) 
&=& 
\frac{N_c^{f} \kappa_{\varphi H}^2}{32 \pi} \frac{s}{(s - m_h^2)^2}  
 \frac{\sqrt{1 - \frac{4 m_\varphi^2}{s}}}{\sqrt{1-\frac{4 m_f^2}{s}}}  
 \left( \frac{m_{f}^2}{s}  \right) 
\left[ 1 - \frac{4 m_{f}^2}{s} \right]
\,,   \label{sigmas} 
\end{eqnarray}
with $N_c^{f}=3(1)$ for quarks (leptons) and 
$\sqrt{s}$ being the center of mass energy.  
Here we have neglected the Higgs width because 
the effective range of $\sqrt{s}$ is much above the Higgs mass scale. 
The number density per entropy density today, $Y(T_0)=n(T_0)/s(T_0)(\simeq Y(T_f))$, 
can be calculated by integrating the Boltzmann equation 
with the production cross sections $\sigma(ij \to \varphi^\dag \varphi)$ 
and the boundary condition $Y(T = m_\varphi \simeq\Lambda_{\rm HC})=0$ 
to be 
\begin{eqnarray}
 Y(T_0) 
&\simeq& Y(T_f) 
\nonumber \\ 
 &=& \frac{135 \sqrt{10} M_p \zeta^2(3)}{32 \pi^7} 
 \int_{T_f}^{\Lambda_{\rm HC}} dT \sum_{i,j} \frac{g_i g_j \eta_i \eta_j}{[g_*(T)]^{3/2}} 
 \int_{(m_i+m_j)/T}^\infty dx 
\nonumber \\ 
&& 
\times 
 x^4 K_1(x) \sigma(ij \to \varphi^\dag \varphi) 
\left( 1 - \frac{(m_i + m_j)^2}{x^2 T^2} \right) 
\left( 1 - \frac{(m_i-m_j)^2}{x^2 T^2} \right)
 \,, \label{YT0}
\end{eqnarray}
where $g_*(T)$ stands for the effective degree of freedom for relativistic particles, 
$g_i=2(1)$ and $\eta_i=3/4(1)$ for fermions (bosons), 
$M_p\simeq 10^{18}$ GeV (reduced Planck mass), $K_1(x)$ denotes the 
modified Bessel function of the first kind, $\zeta(3)\simeq 1.202$, 
$x\equiv \sqrt{s}/T$.

\begin{figure}[t]
\begin{center}
   \includegraphics[width=7.0cm]{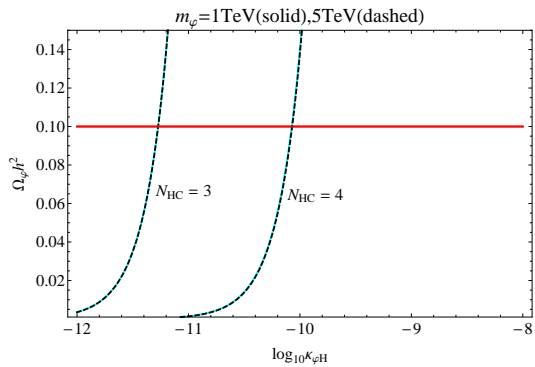} 
\caption{ 
The constraints on the BSP coupling $\kappa_{\varphi H}$ 
from the presently observed relic abundance 
($\simeq 0.1$ drawn as red-horizontal line in the plot) 
for cases of the bosonic baryon $(N_{\rm HC}=4)$ and 
the fermionic baryon ($N_{\rm HC}=3$). 
The plots have been shown by taking the mass to be 1 TeV (solid cyan curves) 
and 5 TeV(dashed black curves). 
\label{kappa-relic}} 
\end{center} 
\end{figure}

The relic abundance, $\Omega_\varphi h^2 = Y(T_0) \cdot m_\varphi s(T_0)/(\rho_{\rm cr}h^{-2})$, 
turns out to actually be almost independent of the $\varphi$ mass 
as far as the $\varphi$ mass of ${\cal O}({\rm TeV})$ is concerned 
(See Fig.~\ref{kappa-relic}). 
The BSP coupling $\kappa_{\varphi H}$ is then constrained 
by the presently observed DM relic abundance $\simeq 0.1$. 
Figure~\ref{kappa-relic} shows the constraint plot on the portal coupling 
for the scalar baryon DM $(N_{\rm HC}=4)$, as well as 
the fermionic baryon case $(N_{\rm HC}=3)$ 
in which $\varphi \sim \psi\psi\psi$~\footnote{ 
In the case of $N_{\rm HC}=3$ the $\varphi \sim \psi\psi\psi$ 
would actually be a spin 3/2 baryon, not be the Dirac fermion of the ground state 
in terms of the spin statistics, though it can be stable by the HC-baryon number conservation.}. 
The figure tells us the upper bounds, 
\begin{eqnarray} 
 \kappa_{\varphi H} 
&\lesssim& 
10^{-10} (10^{-11})
 \,, \nonumber \\ 
 {\rm or} 
\qquad  
 y 
&\lesssim& 
10^{-5} (10^{-6}) \times (1.0/a)^{1/2} 
 \,,\label{upper} 
\end{eqnarray} 
for $N_{\rm HC}=4 (3)$,  
where $g_*(T)$ in Eq.(\ref{YT0}) has been taken to be $\simeq 100$. 
The smallness of the BSP coupling, $\kappa_{\varphi H}$ or $y$ in Eq.(\ref{upper}), 
is consistent with 
the BS mechanism, 
and indeed implies the BSP to be non-thermal. 
Note also that the size of the small $y$ does not affect the 
realization for the Higgs boson mass of 125 GeV: 
after solving the potential problem and fluctuating the 
Higgs field around the EW vacuum, one finds that  
the Higgs mass is given as $m_h \simeq \sqrt{2 \lambda_H} v $ 
with the quartic coupling of the elementary Higgs $\lambda_H$, 
which is precisely the same mass formula as in the SM   
(For details, see  Ref.~\cite{Ishida:2016ogu}).

\section{Conclusion and discussion}

In conclusion, 
the bosonic-seesaw portal scenario, 
based on the model descriptions {\bf (I)} and 
{\bf (II)} proposed in the present paper, 
provides a dark matter candidate having the coupling to the standard model Higgs, 
which is dynamically generated by the seesaw and essentially related to 
the origin of the electroweak symmetry breaking. 
In this scenario the dark matter candidate 
dynamically arises as the hypercolor baryon with 
the conserved hypercolor-baryon charge.  
The composite baryonic dark matter 
can be non-thermally produced enough 
due to the significantly small coupling to the standard model Higgs 
as the consequence of the bosonic seesaw mechanism, to realize 
the observed relic amount which allows for the composite dark matter 
to have the mass of TeV 
scale, in contrast to 
the conventional thermal freeze-out scenario.

Having the Higgs portal coupling,  
the composite baryonic dark matter $\varphi$ can be detected by the 
direct detection experiments such as the LUX~\cite{Akerib:2015rjg}, PandaX-II~\cite{Akerib:2016vxi}, 
and upcoming XENON1T and LZ~\cite{Feng:2014uja}. 
The spin-independent (SI) cross section     
is computed as 
$ 
 \sigma_{SI}(\varphi N \to \varphi N) 
 \simeq \frac{\kappa_{\varphi H}^2}{16 \pi m_h^4} 
m_*^2(N, \varphi) g_{h NN}^2 
$,  
where 
$g_{hNN}\simeq 0.25$ GeV/$v$~\cite{Ohki:2008ff,Oksuzian:2012rzb,Hisano:2015rsa} 
and $m_*(N,\varphi)=m_N m_\varphi/(m_N + m_\varphi)$ is the reduced mass 
with $m_N\simeq 940$ MeV.     
Using the upper bound for the portal coupling $\kappa_{\varphi H}$ in Eq.(\ref{upper}) 
and taking the dark matter mass $1\mathchar`-5$ TeV for the reference value, 
we find the upper bound on the SI cross section, 
$\sigma_{SI} \lesssim 10^{-63} (10^{-65}) \, {\rm cm}^2$ 
for $N_{\rm HC}=4(3)$. 
These values are far below the current limit most stringently set by the 
LUX2016~\cite{Akerib:2015rjg}, 
and the sensitivity in the future-prospected XENON1T or LZ, 
$\sigma_{SI} \le 10^{-47} \,{\rm cm}^2$ at the TeV range~\cite{Feng:2014uja}, which will 
actually be overlapped with the expected neutrino background~\cite{Billard:2013qya}.

Thus, since having the extremely small portal coupling,  
the bosonic-seesaw portal dark matter is fairly insensitive to the direct 
detection experiments, which would imply some extension for the present 
scenario, or other detection proposals.

\acknowledgments 
We are grateful to Takehiko Asaka for valuable comments. 
This work was supported in part by 
the JSPS Grant-in-Aid for Young Scientists (B) \#15K17645 (S.M.) 
and Research Fellowships of the Japan Society for the Promotion of Science for Young Scientists
 \#26$\cdot$2428 (Y.Y.).

\end{document}